\documentclass{elsart}
\usepackage{natbib}
\usepackage{amssymb,amsmath}
\usepackage{graphicx}

\newcommand{\be}  { \begin{equation} }
\newcommand{\ee}  { \end{equation}   }
\newcommand{\bea} { \begin{eqnarray} }
\newcommand{\eea} { \end{eqnarray}   }

\newcommand{\msun}{ \ensuremath{M_{\odot}  } }

\newcommand{\pdot}{ \ensuremath{\dot P     } }
\newcommand{\pddot}{ \ensuremath{\ddot P     } }
\newcommand{\rhogj}{ \ensuremath{\rho_\mathrm{GJ} } }
\newcommand{\Edot}{ \ensuremath{\dot E     } }
\newcommand{\edot}{ \ensuremath{\dot E     } }
\newcommand{\Idot}{ \ensuremath{\dot I     } }
\newcommand{\Ic}{ \ensuremath{I} }
\newcommand{\rmd}{ \ensuremath{{\rm d}} }
\newcommand{\Omegadot}{ \ensuremath{\dot \Omega   } }

\begin{document}
\begin{frontmatter}
\title{What can the braking indices tell us about pulsars' nature?}

\author{Y. L. Yue\corauthref{cor1}},
\corauth[cor1]{Corresponding author.} \ead{yueyl@bac.pku.edu.cn}
\author{R. X. Xu},
\author{W. W. Zhu}
\address{Astronomy Department, School of Physics, Peking University,
         Beijing 100871, China}

\begin{abstract}
As a result of observational difficulties, braking indices of only
six rotation-powered pulsars are obtained with certainty, all of
which are remarkably smaller than the value ($n=3$) expected for
pure magnetodipole radiation model. This is still a real
fundamental question not being well answered after nearly forty
years of the discovery of pulsar. The main problem is that we are
shamefully not sure about the dominant mechanisms that result in
pulsars' spin-down. Based on the previous works, the braking index
is re-examined, with a conclusion of suggesting a constant gap
potential drop for pulsars with magnetospheric activities. New
constrains on model parameters from observed braking indices are
presented.
%
\end{abstract}


\end{frontmatter}

\section{Introduction}
\label{}

In pulsar emission models, we usually have the assumption \be
\dot{\Omega} \propto \Omega^{n}, \ee where $\Omega = 2\pi/P$, $P$
is the spin period, the index, $n$, is usually assumed to be a
constant measuring the efficiency of braking.
We can then define the braking index, $n$, and the second braking
index, $m$, \be n = \frac{\Omega \ddot{\Omega}}{\dot{\Omega}^2},
\quad m = \frac{\Omega^2 \dddot{\Omega}}{\dot{\Omega}^3}. \ee
By investigating $n$ and $m$, we can get much information of
pulsar's radiation and spin-down mechanism. In those models where
spindown due to pure magnetodipole radiation is assumed, one has
only $n=3$.

The fairly accurate timing property of pulsars give us the
opportunity to measure not only the period $P$ and the period
derivative $\pdot$, but also the second period derivative
$\ddot{P}$ and even the third.
However, there do exist difficulties in observation. As a result,
braking indices of only six rotation-powered pulsars are obtained
with certainty, i.e. PSR J1846-0258   [$n=2.65(1)$], PSR
B0531$+$21   [2.51(1), the Crab pulsar], PSR B1509$-$58
[2.839(3)], PSR J1119$-$6127 [2.91(5)], PSR B0540$-$69 [2.140(9)]
and PSR B0833$-$45   [1.4(2), the Vela pulsar], where the digits in
the parentheses indicate the uncertainties of the last digits
\citep[][and references therein]{Livingstone06}. All these six
braking indices are smaller than the value ($n=3$) predicted by
pure dipole magnetic field configuration, which may suggest that
other spin-down torques do work besides the energy loss via dipole
radiation.

Pulsar's spin-down has been studied since its discovery, but why
the braking indices are smaller than 3 still does not have a clear
answer yet. In recent years, several new mechanisms are suggested to
explain this discrepancy, e.g., the two-component model: spin-down
due to both magnetodipole radiation and relativistic particle flow
\citep{xuqiao01,Wu03,Contopoulos06}, the models with changing
inclination angles \citep[e.g.][]{Ruderman05}, the models with
increasing magnetic field strength
\citep[e.g.][]{linzhang04,Lyne04,Chen06}, the
models with additional torques due to accretion
\citep[e.g.][]{Chen06} and the model with field-reconnection in
magnetosphere \citep[e.g.][]{Contopoulos06recon}.

In this paper, we focus on the two-component model suggested by
\citet{xuqiao01}. Assuming a constant potential drop
($\sim10^{12}$ V) in the polar cap accelerating region, we have
only one free parameter left. Comparing with observations, new
constrains are presented. The six pulsars with measured braking
indices can be divided into two groups. The charge density in the
polar cap region could be $\sim10^3$ times the the
Goldreich-Julian charge density $\rhogj$ \citep{GJ69}, much larger
than the usually assumed value $\sim\rhogj$. The derived magnetic
field strength could be larger than the value that from pure
magnetodipole assumption. Additionally, the change of momentum of
inertia $I$ would also affect the braking index and make it
smaller than 3.

\section{The model with constant gap potential drop}

In the two-component models,
the magnetic momentum is assumed to be
$\boldsymbol{\mu}=\boldsymbol{\mu}_\bot + \boldsymbol{\mu}_\parallel$,
where $\mu = BR^3/2$, $\mu_\bot = \mu\sin\alpha$,
 $\mu_{\parallel} = \mu\cos\alpha$
\citep{xuqiao01, Contopoulos06} and $\alpha$ is the inclination
angle. The component $\mu_\bot$ produces magnetodipole radiation,
while the other component $\mu_{\parallel}$ could have various
choices of acceleration mechanisms such as inner vacuum gap with
curvature radiation \citep[VG, e.g.][hereafter RS75]{RS75}, vacuum
gap with resonant inverse Compton scattering \citep[VG+ICS,
e.g.][]{Zhang00}, outer gap \citep[OG, e.g.][]{Cheng86} and space
charge limited flow \citep[SCLF, e.g.][]{Arons79}.

The energy loss rate $\edot$ of the two components usually have
different dependencies on $\Omega$. The magnetodipole component
has $\edot_\bot \propto\Omega^4$, which would induce a braking
index of 3. The other component usually has a relatively weaker
dependence on $\Omega$ (e.g. $\edot_\parallel \propto \Omega^{2}$
for  a constant gap potential drop due to the unipolar effect),
which would induce a braking index less than 3 ($n=1$ in that
case). The magnetodipole radiation is dominant when $P$ is shorter
while the other component becomes important when $P$ is longer.
The combination of the two would explain the observed braking
index between 1 and 3. In summary, there are three
variables/parameters: $B$, $P$ and $\alpha$.

Here we firstly use the RS75 inner vacuum gap model for indication.
The other models which can also be parameterized the same way will
be discussed in \S \ref{discuss}.

The energy lose rates of dipole and unipolar are
$\edot_\mathrm{dip} = -(2/3)c^{-3}\mu^2 \Omega^4 \sin^2\alpha$
and
$\edot_\mathrm{uni} = -2\pi r_\mathrm{pc}^2 \rho\Phi
= -c^{-1} \kappa BR^3 \Omega^2 \Phi \cos^2\alpha$,
where $r_\mathrm{pc}$ is the polar cap radius,
$\rho=\kappa\rhogj$ is the charge density of the par cap region,
$\rho_\mathrm{GJ} = B/(cP)$ is the Goldreich-Julian
charge density \citep{GJ69} and $\kappa$ is an uncertain coefficient.
We use $\kappa$ as another free parameter,
which would be constrained by the observations.
Combining these two, we have
\be
\edot = I \Omega \dot{\Omega} = -\frac{2}{3c^3}\mu^2 \Omega^4 \eta,
\label{eq:edotdip}
\ee
where
\be
\eta = \sin^2\alpha +
6c^2 \kappa B^{-1} R^{-3} \Omega^{-2} \Phi \cos^2\alpha.
\ee
The effective potential drop $\Phi$ of unipolar usually has a weak
dependence on $\Omega$ \citep[e.g. $\Phi \sim \Omega^{-1/7}$ in][]{xuqiao01},
or just a few$\times10^{12} \mathrm{~V}$ \citep[e.g. RS75;][]{Usov95},
so we could assume a constant potential drop
$\Phi=10^{12}$ V in the polar gap region.
At the same time, a potential drop of $10^{12}$ V is also
a widely-accepted result from pulsar death-line criterion.
The braking index and the second braking index then are
\be
n = 3 - 2\frac{\Omega^{-2}}{\tan^2\alpha/f + \Omega^{-2}}
\label{eq:n}
\ee
and
\be
m = 2 n^2 -n + (n-3)(1-n),
\label{eq:m}
\ee
where $f = 6c^2 \kappa B^{-1} R^{-3} \Phi$.

\begin{figure}
\begin{center}
\includegraphics*[width=.5\textwidth]{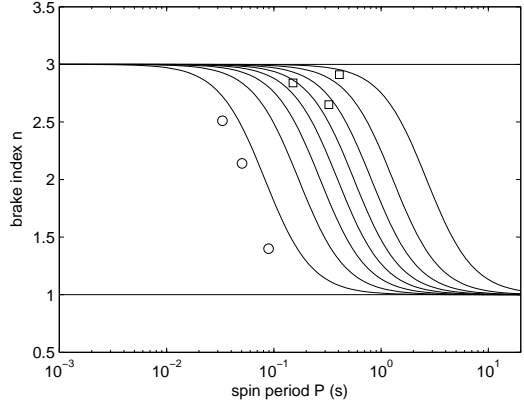}
\end{center}
\caption{
Braking index $n$ versus spin period $P$ for the two-component model.
Here we use magnetic dipole plus unipolar with constant potential drop
$\Phi=10^{12}\mathrm{~V}$.
The magnetic field is assumed to be $B=10^{12} \mathrm{~G}$
and the charge density is assumed to be $\rho=10^3\rhogj$.
The lines from bottom to top correspond to $\alpha$ from
$0^\circ$ to $90^\circ$ in $10^\circ$ increments.
The six pulsars could be divided into two groups:
three pulsars (squares, Group I) have larger spin period $P$,
larger braking indices
and larger $\alpha$
while the other three (circles, Group II) have relatively
smaller $P$, smaller braking indices
and smaller $\alpha$.
There is a gap between these two groups.
It should be noted that we could not get $\alpha$ directly from this graph,
because $B$ is not all equals to $10^{12} \mathrm{~G}$
and $\rho$ is a parameter.
This figure is for indication.
\label{fig:1}}
\end{figure}

Assuming  a pulsar of $1.4\msun$ in mass and 10 km in radius with known
$P$, $\pdot$, $\pddot$ and a constant potential drop $\Phi=10^{12}$ V,
 we have three variable: $B$, $\alpha$ and $\rho$.
Considering two constrains from Eqs. (\ref{eq:edotdip}) and (\ref{eq:n}),
we have only one free parameter.
Here we use $\rho$ as the free parameter
and solve out $B$ and $\alpha$,
i.e. $B$ and $\alpha$ are plotted as function of $\rho$.
Our results are presented in Figs. 1--4.
The six pulsar could be divided into two groups (see Fig. [\ref{fig:1}]):
three that has larger spin period $P$ and larger braking indices $n$,
i.e. PSR J1846$-$0258, PSR B1509$-$58 and PSR J1119$-$6127 (Group I);
three that has smaller $P$ and smaller $n$,
i.e. Crab, Vela and PSR B0540$-$69 (Group II).
The charge density $\rho$ could be around $10^3\rhogj$
to make $B$ and $\alpha$ have reasonable values (Fig. [\ref{fig:3}]).
If the braking index $n$ is close to 3,
the spin down is mainly due to magnetodipole radiation.
The $B$ value then is close that from the canonical formula
$B_\mathrm{dip} = 6.4\times10^{19} (P\pdot)^{1/2}$,
i.e. the dipole approximation is quite good.
If the braking index $n$ is close to 1,
the spin down torque is mainly from particle outflow.
The $B$ value would depart from $B_\mathrm{dip}$
considerably (Fig. [\ref{fig:4}]).

\section{The change of momentum of inertia}

In a more general case, we have \be \Edot = I\Omega\Omegadot
+\frac{1}{2}\Idot\Omega^2 \label{eq:edot} \ee In \S2, we only
consider the first term $I\Omega\Omegadot$ in Eq. (\ref{eq:edot})
and omitted the second term $\Idot\Omega^2/2$, i.e. we approximate
$I$ as a constant. In this section we calculate the effects of
$\Idot$ in two case: (i)  volume conservative, i.e. the pulsar's
volume is a constant, (ii) volume non-conservative, i.e. assume
that the pulsar's volume changes, or equivalently the stellar
radius $R$ changes. W show that the change of braking index  is
quite small in the first case but should be considered in the
second one.

\subsection{The volume conservative case}

A rotating star, or specifically a pulsar, can be approximately
as an Maclaurin rotation ellipsoid
when the spin period is not too small \citep{Zhou04}.
For pulsars, the criteria is $P \gg 1$ ms,
so the rotation ellipsoid approximation are available
for all the six pulsar in this article.
In this case the pulsar's volume is assumed to be a constant,
i.e. the pulsar is incompressible.
The pulsar is a spherical when $\Omega=0$ and is a rotation ellipsoid when $\Omega > 0$.
We have \citep{Zhou04}
\be
I = I_0(1+\frac{1}{3}e^2),
\ee
where
\be
e = \frac{\Omega}{ \sqrt{8\pi G \rho_*/15} },
\ee
$\rho_*$ is the average star density and $I_0$ is
the star's momentum of inertia when $\Omega=0$.
For Crab, the fastest one in the six, $e=0.022$.
Here we define
\be
I = I_0(1+k\Omega^2),
\label{eq:I}
\ee
where $k = 5/(8\pi G \rho)$.
Then we have
\be
\Idot = \frac{{\rm d}I}{ {\rm d}\Omega } \frac{{\rm d}\Omega}{ {\rm d}t }
=2kI_0\Omega\Omegadot
\label{eq:Idot}
\ee
Assuming $\edot$ is a power-law function of $\Omega$, we have
\be
I\Omega\Omegadot +\frac{1}{2}\Idot\Omega^2 = k_u\Omega^{u+1},
\label{eq:upara}
\ee
where $u$ is a constant (usually between 1 and 3) and $k_u$ is a coefficient.
Here $u$ equals to braking index
if we omit the $\Idot\Omega^2/2$ term.
With Eqs. (\ref{eq:I}), (\ref{eq:Idot}) and (\ref{eq:upara}),
we have
\be
\Omegadot = \frac{k_u}{I_0}\frac{\Omega^{u}}{1+2k\Omega^2}
\ee
and
\be
\ddot{\Omega} = \frac{k_u}{I_0}\frac{\Omega^{u-1}\Omegadot}{1+2k\Omega^2}
(u-\frac{4k\Omega^2}{1+2k\Omega^2}).
\ee
Then we have the braking index
\be
n' = \frac{\ddot{\Omega}\Omega}{\Omegadot^2}
=u-\frac{4k\Omega^2}{1+2k\Omega^2}.
\label{eq:nnew}
\ee
Or effectively we have the difference between
the braking index when assuming a constant $I$ and
the braking index when considering $\Idot$,
\be
\Delta n = n -n'= u-n' = \frac{4k\Omega^2}{1+2k\Omega^2}.
\ee
For Crab,
$k\Omega^2=e^2/3 = 1.6\times10^{-4} \ll 1$,
$\Delta n =6.5\times10^{-4}$,
which could be omitted.
Since Crab is the fastest one of the six, this effect
can be omitted for all the six pulsars.

\subsection{The volume non-conservative case}

In the above subsection we show that the volume conservative case
does not have much effects on braking index, because the change of
$I$ is very small. Here we consider a more effective way, in which
$\Idot$ is larger, i.e. the star radius $R$ changes. Here we
assume that $I$ is a function of $\Omega$. Applying the first
order approximation, we have $\rmd I/\rmd \Omega$ = const. Here we
define \be \frac{\rmd I}{\rmd \Omega} = \Ic f_1 \ee where \Ic is
the current moment of inertia and $f_1$ is a coefficient. Together
with Eq. (\ref{eq:upara}), we have \be \Omegadot = \frac{k_u }{ I
}\frac{\Omega^u}{1+f_1\Omega} \ee and then \be \ddot{\Omega} =
\frac{k_u }{ I }\frac{\Omega^{u-1}\Omegadot }{ 1+f_1\Omega}
(u-\frac{f_1\Omega}{1+f_1\Omega}). \ee The braking index is \be n
= u-\frac{f_1\Omega}{1+f_1\Omega}. \ee The value of $f_1\Omega$
should be of the order of 1 to make braking index be obviously
smaller than $u$ ($u$ is the braking index if $\Idot=0$  is
assumed). For Crab we have, \be \Idot = \frac{\rmd I}{\rmd \Omega}
\Omegadot =I f_1 \Omegadot \sim I \frac{1}{\Omega} \Omegadot = 4
\times10^{-4}\Ic {\rm ~yr^{-1}}. \ee It means the $I$ of Crab
changes $\sim1\%$ in the past 30 years. This seems large but is
possible. Actually, a prontoneutron star could have a radius $\sim
30$ km but a cooled one may have radius $\sim 10$ km
\citep{Lattimer04}. The momentum of inertia changes dramatically.
The six pulsars which have measured braking indices are all young
and may be still cooling. In addition, strain energy increases as
a solid star spins down, which may also cause the decreasing of
stellar volume \citep{xty06}. It is therefore not unreasonable if
they have such values of $\Idot$. It is worth noting that the
``observed'' field increasing \citep[e.g. ][]{linzhang04} of the
Crab pulsar could also arise from the shrinking of pulsars after
quakes.

\section{Conclusion and discussion}
\label{discuss}

We present a calculation of pulsar's braking index considering
magnetodipole radiation plus RS75 inner vacuum gap with a constant
potential drop. The six pulsars with measured braking index tends
to have either small or big $\alpha$, i.e. they can be divided
into two groups and there is a gap between the two group.
Assumed a constant potential drop $\Phi = 10^{12}\mathrm{~V}$, we
get that the charge density $\rho$ in the polar cap region is
$\sim 10^3$ times of $\rhogj$, much larger the usually assumed
value $\sim\rhogj$. The dynamical implication of this assumption
in pulsar electrodynamics is worth to research in the future.
We also shows that the effects of $\Idot$ on braking index
could not be omitted if $\Idot$ is of the order of $10^{-4}I\mathrm{~yr^{-1}}$.

In our results, the value of $\rho\sim 10^3\rhogj$ is much larger
than $\rhogj$. This means that $e^\pm$ pair plasma could be
accelerated in gaps with potential drop $\Phi$. Though RS75-type
vacuum gap may result in sparking, which would be necessary for
explaining drifting subpulses, it is still possible that pair
plasma could be accelerated (i) above the vacuum gap and/or (ii)
in the annular gap in \citet{Qiao04}.
If not only VG but also ICS and/or OG exists
\citep[e.g. VG+ICS model in][]{Zhang00},
i.e. the effective
potential drop is larger than the potential drop of the inner gap,
the charge density could be smaller.
Although a constant potential drop of $10^{12}$ V is generally accepted,
we would like to note that $\Phi$ might changes slightly with $\Omega$
\citep[e.g. $\Phi \sim \Omega^{-1/7}$ in][]{xuqiao01}.
The effects would change the braking index by a factor, e.g. $\Delta n\sim-1/7$ if
$\Phi \sim \Omega^{-1/7}$.

The second braking indices of two pulsars are measured,
i.e. Crab ($m=10.23\pm0.03$) and PSR B1509-58 ($m=18.3\pm2.9$) \citep{Livingstone06}.
The theoretical value form Eq. (\ref{eq:m}) is $m=10.9$ for Crab
and $m=13.6$ for  PSR B1509-58,
which are not compatible with the observations within the error bar.
However, this discrepancy might not be a serious problem,
because the braking indices are affected by several processes such as
very small glitches and changing of surrounding environments (e.g. interaction with ISM),
which is not clearly understood.

Are there really two groups of pulsars? Is there really a gap?
The answer could not be certain yet because
we only have braking index from six pulsars.
There might be some mechanism to affect the inclination angle
especially while the pulsar forms in process of the supernova explosion,
but this is not well understood.
When the number of pulsars with measured braking indices
doubled in the future, the answer
to this question would be much more clear.

\section*{acknowledgements}
We would like to thank Prof. Guojun Qiao for his helpful
suggestions and to acknowledge various stimulating discussions in
the pulsar group of Peking university.
This work is supported by
National Nature Science Foundation of China (10573002) and by the
Key Grant Project of Chinese Ministry of Education (305001).

\begin{figure}
\begin{center}
\includegraphics*[width=.5\textwidth]{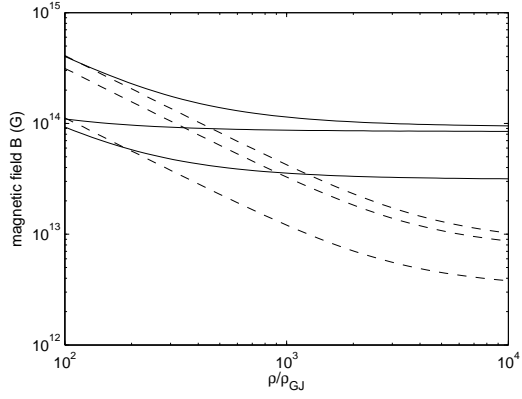}
\end{center}
\caption{
Magnetic field versus normalized charge density $\rho/\rhogj (=\kappa)$.
The $B$ value of pulsars with larger $P$ (solid line, Group I)
do not change much,
but it changes substantially for
the pulsars with smaller $P$ (dashed line, Group II).
\label{fig:2}}
\end{figure}

\begin{figure}
\begin{center}
\includegraphics*[width=.5\textwidth]{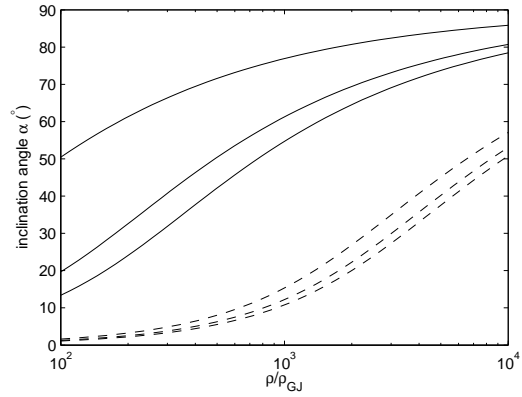}
\end{center}
\caption{
Inclination angle $\alpha$ versus normalized charge density $\rho/\rhogj (=\kappa)$.
Solid line for the three pulsar with larger $P$ (Group I),
dashed line for the three pulsars with smaller $P$ (Group II).
There is a clear gap between this two groups.
If we assume that $\alpha$ is neither too big nor too small,
we have $\rho$ around $10^{3} \rhogj$.
\label{fig:3}}
\end{figure}

\begin{figure}
\begin{center}
\includegraphics*[width=.5\textwidth]{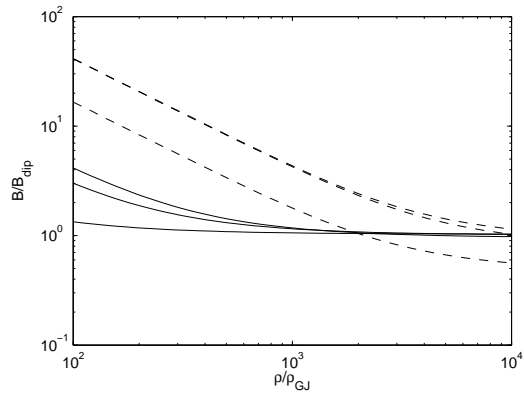}
\end{center}
\caption{
$B/B_\mathrm{dip}$ versus normalized charge density $\rho/\rhogj (=\kappa)$.
Solid line for the three pulsar with larger $P$ (Group I), dashed line for
the three pulsars with smaller $P$ (Group II).
If we assume $\rho$ is around $10^{3}\rhogj$  (see Fig. \ref{fig:3}),
the $B$ values of Group I pulsar is close to
$B_\mathrm{dip}$ with in a factor of 2.
The $B$ value of Group II pulsar in this case
is a few times the value of $B_\mathrm{dip}$.
\label{fig:4}}
\end{figure}

\end{document}